 \documentclass[final,3p,times]{elsarticle}
\usepackage{amssymb}
\usepackage[fleqn]{amsmath}
\journal{arXiv}

\begin{document}

\begin{frontmatter}

%% Title, authors and addresses

%% use the tnoteref command within \title for footnotes;
%% use the tnotetext command for theassociated footnote;
%% use the fnref command within \author or \address for footnotes;
%% use the fntext command for theassociated footnote;
%% use the corref command within \author for corresponding author footnotes;
%% use the cortext command for theassociated footnote;
%% use the ead command for the email address,
%% and the form \ead[url] for the home page:
%% \title{Title\tnoteref{label1}}
%% \tnotetext[label1]{}
%% \author{Name\corref{cor1}\fnref{label2}}
%% \ead{email address}
%% \ead[url]{home page}
%% \fntext[label2]{}
%% \cortext[cor1]{}
%% \address{Address\fnref{label3}}
%% \fntext[label3]{}

%%%%%%%%%%%%%%%%%%%%%%%%%%%%%%%%%%%%%%%%%%%%%%%%%%%%%%%%%%%%%%%%%%%%%%%%%%%%%%
%%%%%%%%%%%%%%%%%%%%%%%%%%%%%%%%%%%%%%%%%%%%%%%%%%%%%%%%%%%%%%%%%%%%%%%%%%%%%%

% \title{Effect of bending in unentangled dilute polymer dynamics revisited}

\title{Time-correlated forces and biological variability in cell motility}

%% use optional labels to link authors explicitly to addresses:
%% \author[label1,label2]{}
%% \address[label1]{}
%% \address[label2]{}

\author{T. N. Azevedo$^{1}$} %

%\author{M. L. Martins$^{1,\ddagger}$} %
%\ead{mmartins@ufv.br}

\author{L. G. Rizzi$^{1,*}$} %
%\ead{lerizzi@ufv.br}

\address{$^{1}$\,Departamento de F\'isica, Universidade Federal do Vi\c{c}osa, CEP~36570-000, Vi\c{c}osa, MG, Brazil.}

%\address{$^{\ddagger}$\,Instituto de F\'isica de Ibitipoca, IBITIFIS, CEP~36140-000, Concei\c{c}\~ao do Ibitipoca, MG, Brazil.}

\date{\today}

%%%%%%%%%%%%%%%%%%%%%%%%%%%%%%%%%%%%%%%%%%%%%%%%%%%%%%%%%%%%%%%%%%%%%%%%%%%%%%
%%%%%%%%%%%%%%%%%%%%%%%%%%%%%%%%%%%%%%%%%%%%%%%%%%%%%%%%%%%%%%%%%%%%%%%%%%%%%%

\begin{abstract}
\noindent
	Cell motility is one of the most fundamental phenomena underlying biological processes that maintain living organisms alive.
      Here we introduce a simple model to describe the motility of cells which include not only time-correlated internal forces but also the biological variability which is inherent of the intra-cellular biochemical processes.
	Such model allow us to derive exact expressions for the mean-squared displacement and the effective time-dependent diffusion coefficient which are compared to numerical results obtained from non-markovian stochastic simulations.
	In addition, we show that the heterogeneity of persistence times lead to non-gaussian distributions which can be obtained analytically and that were validated by the numerical simulations.
	Our results indicate that such model might be used to describe the behaviour observed in experimental results obtained for isolated cells without external signaling.
\end{abstract}

%%%%%%%%%%%%%%%%%%%%%%%%%%%%%%%%%%%%%%%%%%%%%%%%%%%%%%%%%%%%%%%%%%%%%%%%%%%%%%
%%%%%%%%%%%%%%%%%%%%%%%%%%%%%%%%%%%%%%%%%%%%%%%%%%%%%%%%%%%%%%%%%%%%%%%%%%%%%%

\begin{keyword}
%% keywords here, in the form: keyword \sep keyword
%% PACS codes here, in the form: \PACS code \sep code
%% MSC codes here, in the form: \MSC code \sep code
%% or \MSC[2008] code \sep code (2000 is the default)

cell motility \sep time-correlated forces \sep biological variability %\sep active diffusion

\end{keyword}

\end{frontmatter}

%% \linenumbers

%%%%%%%%%%%%%%%%%%%%
%%% INTRODUCTION %%%
%%%%%%%%%%%%%%%%%%%%

\section{Introduction}

      Active physical and chemical processes that affect the cellular behaviour
%are inherent to
play an important role in the homeostasis of almost all living 
organisms~\cite{hofling2013repprogphys}.
      Cell motility, in particular, % in particular, % is an important aspect which 
influences numerous fundamental biological processes, such as organogenesis ({\it i.e.},~cell 
aggregation and migration), wound healing, and tumor development~\cite{petrie2009natrev,mierkefabry2008eurjcellbiol,weiger2013plosone}.

\begin{figure}[!b]
\centering
\includegraphics[width=0.45\textwidth]{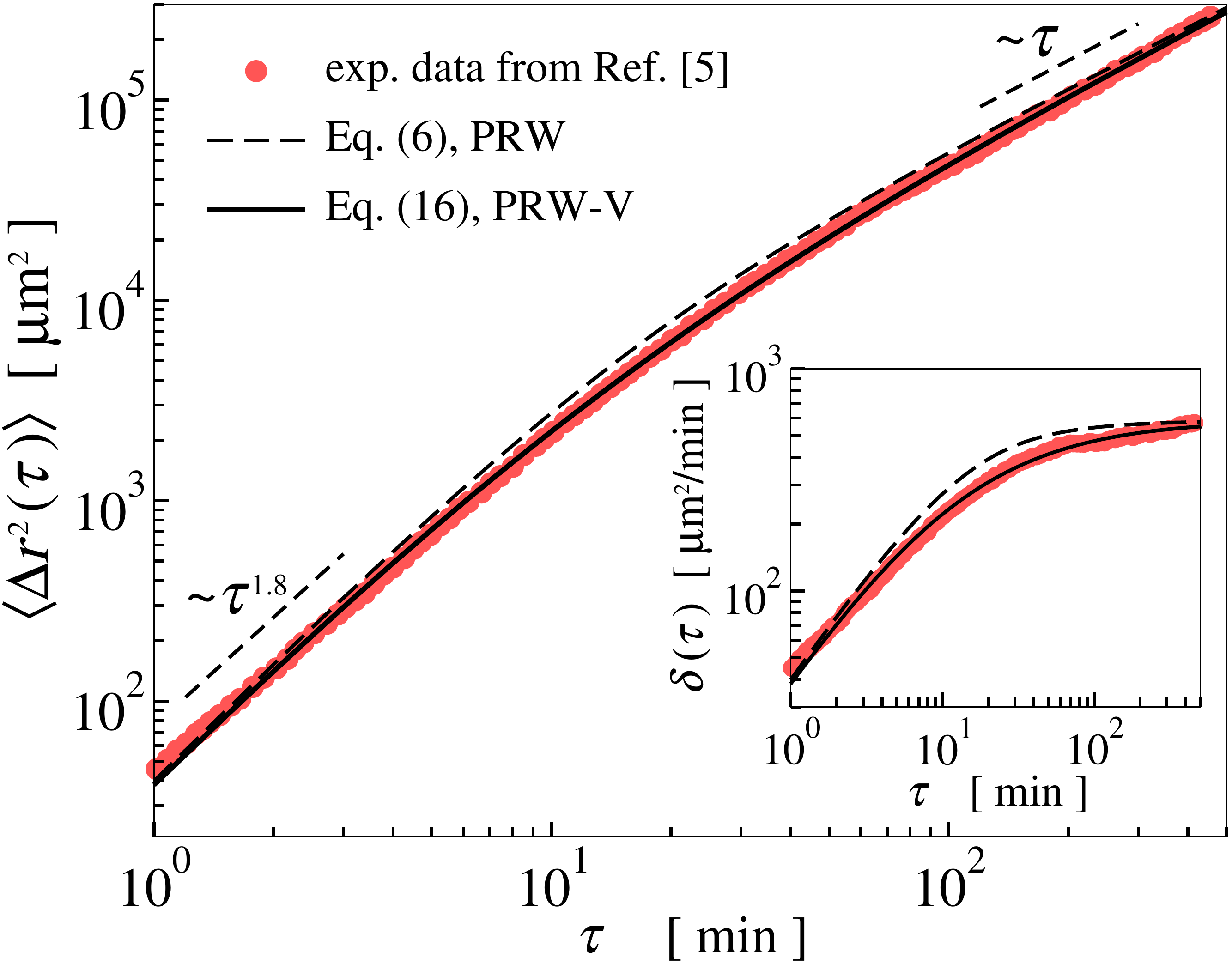}
\caption{Comparison between the experimental data extracted from Ref.~\cite{cox2008plosone} (denoted by the filled red circles)
and the theoretical results obtained from 
the usual persistent random walk (PRW) model, and
the persistent random walk model with variability (PRW-V) introduced here.
The main panel display the mean-squared displacement $\langle \Delta r^2 (\tau) \rangle$,
while the inset shows the ratio $\delta(\tau) = \langle \Delta r^2 (\tau) \rangle / \tau$.
Short dashed lines indicate the short-time superdiffusive, or ballistic, ($\alpha>1$) and
the normal diffusive behaviours ($\alpha\approx 1$) observed at later times.
Long dashed and straight lines correspond to the PRW model, Eq.~(\ref{PRW-MSD}), and
the PRW-V model, Eq.~(\ref{PRW-V-MSD}), respectively.
Both models are described by the velocity autocorrelation parameter $v_0^{\,}=4.57\,\mu$m/min 
and a persistence time $\tau_{p}=7\,$min.
}
\label{exp_data_isolated_cell}
\end{figure}

	In general, the stochastic dynamics of the cell movement exhibit properties that differ from the normal diffusive-like behaviour,
%and often reveal characteristics of anomalous diffusion~\cite{metzler2001physrep}. 
%      This deviation
which is often characterized by a mean-squared displacement that is given by
\begin{equation}
\langle \Delta r^2(\tau) \rangle \propto \tau^{\alpha} ~~,
\label{anomdif}
\end{equation}
where the exponent $\alpha$ gives a measure wether their diffusive-like movement is anomalous ({\it i.e.}, when $\alpha \neq 1$)
or not.
% or subdiffusive when $\alpha < 1$.
	As illustrated in Fig.~\ref{exp_data_isolated_cell}, the experimental data extracted from Ref.~\cite{cox2008plosone}  obtained from the motion of isolated cells ({\it i.e.}, without external signaling processes) provide an important evidence that cell's motility can be characterized by a superdiffusive behaviour at short times (with $\alpha > 1$) and by a normal diffusion regime (with $\alpha \approx 1$) at later times.

%	 experiments on motion of isolated cells (in the absence of external signaling), presented in Ref.~\cite{cox2008plosone}, and which is reproduced in Fig.~\ref{exp_data_isolated_cell}.
%	As one can see, the experimental data display a short time superdiffusive behaviour ($\alpha\approx 1.8$) for short times and an almost normal diffusion ($\alpha \approx 1$) at later times, which is in fair agreement with the results from both our simulations and theoretical approach.

%==================

	Although several models have been proposed to describe the different diffusive regimes observed in cell motility~\cite{selmeczi2005biophysj,peruani2007prl,dieterich2008pnas,campos2010jtheorbiol,safaeifard2018review,ritaalmeida2020physicaA}, 
the persistent random walk (PRW) model is one of the most used models to describe experimental data
(see, {\it e.g.},~refs.~\cite{cox2008plosone,wirtz2014pnas,luzhansky2018aplbioeng}).
	Besides its simplicity, such model captures the main feature of cell motility which is the
time-correlation ({\it i.e.},~the persistence) of the random walk~\cite{petrie2009natrev}.
	As shown in Fig.~\ref{exp_data_isolated_cell}, the PRW model can be used to describe the
experimental data for the MSD quite well.~However, 
	a closer look at the experimentally observed displacements and velocities distributions (see, {\it e.g.},~\cite{ueda2008plosone,ebata2018scirep,huda2018natcommun}) reveals that they might present non-gaussian tails which cannot be
satisfactorily described by the usual PRW model.
%	In addition, more complicated models which include orientational correlation
%	Importantly, theoretical developments have showed that orientational correlation could lead to complex transient behaviour between superdiffusive and normal diffusive regimes~\cite{peruani2007prl}.
%Cell heterogeneity~\cite{wirtz2014pnas}
%Cell-to-cell variability~\cite{broedersz2020jrsocinter}
%Cellular mechanisms~\cite{petrie2009natrev}
	Interestingly, there are recent experimental evidence~\cite{petrie2009natrev,wirtz2014pnas,broedersz2020jrsocinter} suggesting that a possible source of the non-gaussianity of the distributions can be the inherent heterogeneities of 
intra-cellular processes that are present even in the absence of external signals.
% the cell-to-cell variability, and 
% aging effects as well as 
%cell-to-cell variability 
%should be important to describe their motility even in the absence of external signals.

	Here we present a generalization of the PRW model which incorporates the cell-to-cell variability by considering that the persistence times $\lambda$ of the internal processes which occur in each cell are distributed according to a distribution $\rho(\lambda)$, and show how it can alter the distributions of distances and velocities.

%%%%%%%%%%%%%%%%%%%%%%%%%%%%%%%%%%%%%%%%%%%%%%%%%%%%%%%%%%%%%%%%%
%%%%%%%%%%%%%%%%%%%%%%%%%%%%%%%%%%%%%%%%%%%%%%%%%%%%%%%%%%%%%%%%%

%%%%%%%%%%%%%%%%%%%%%%%%%%%%%
%%% MODEL AND SIMULATIONS %%%
%%%%%%%%%%%%%%%%%%%%%%%%%%%%%

\section{Time-correlated forces and stochastic simulations}

	Before introduce our model, we present some of the theoretical aspects related to the 
stochastics processes that characterize the PRW model as well as 
the numerical methodology which can be used to obtain its non-markovian dynamics.

%	Since our theoretical and computational approaches share the same stochastic 
%description, we illustrate them by considering a some analytical results derived from the PRW model
%as well as the numerical methods used to obtain its non-markovian dynamics.

%and our model, which we termed {\it persistent random walk with variability}, or PRW-V, 
%and the numerical methods used to obtain 
%related to it.

\subsection{Langevin approach}

	Consider that the cell movement can be effectively described by a simple force equation that is written as an overdamped Langevin-like equation as
\begin{equation}
\gamma \frac{d \vec{r}}{dt} = \vec{f}_s ~~,
\label{eq1}
\end{equation}
%\begin{equation}
%m \frac{d \vec{v}}{dt} = - \gamma \vec{v} + \vec{f}_s,
%\end{equation}
where $\gamma$ is an effective adhesion (or friction) coefficient between the cell and the 
substrate, %adhesion
and $\vec{f}_s$ is a stochastic intra-cellular (or internal) force~\cite{langefabry2013expcellres}.
	Thus, the velocity of the cell, $\vec{v}=\gamma^{-1} \vec{f}_{s}$, %of the cell 
can be also considered a stochastic variable.~From
	the statistical point of view, we assume that it
has the following properties: (i)~$\langle \vec{v}\, \rangle = \vec{0}\,\mu$m/min ({\it i.e.}, zero mean value); and, (ii)~an autocorrelation that is given by
\begin{equation}
\langle \vec{v}(t) \cdot \vec{v}(t') \rangle = d v_0^2 \, A(t-t')~~,
\label{Vautocorr}
\end{equation}
where $v_0^{\,}\equiv f_{s}/\gamma$ is the velocity autocorrelation parameter, %\equiv f_{s}/\gamma$, 
$d$ is the dimension of the substrate, 
and $A(t-t')$ is an autocorrelation function that will be specified later.~Here
	the average $\langle \dots \rangle$ is considered to be taken over an {\it ensemble} of independent cells.
	Accordingly, in order to incorporate the variability of those cells into our theoretical framework, 
we consider that the autocorrelation function can be written as %in terms of a distribution of persistence times as
\begin{equation}
A(\tau) = \int_0^{\infty} \rho(\lambda) \, e^{-\tau/\lambda} \, d\lambda~~,
\label{autocorr_general}
\end{equation}
where $\tau=t-t'$, and $\rho(\lambda)$ is the distribution of persistence times $\lambda$.

%%%%%%%%%%%%%%%%%%%%%%%%%%%%%%%%%%%%%%%%%%%%%%%%%%%%%%%%%%%%

\subsection{Persistent random walk (PRW) model}
\label{PRWmodel}

	The PRW model can be easily retrieved from the above framework by considering that 
there is only a single persistence time $\tau_{p}$ which is common to all cells.~This 
	is done by choosing a delta distribution, {\it i.e.},~$\rho(\lambda)=\delta(\lambda - \tau_{p})$, so that 
Eq.~(\ref{autocorr_general}) yields
$A(\tau) = e^{-\tau/\tau_{p}}$, % where $\tau_{p}$ is the persistence time.
and Eq.~(\ref{Vautocorr}) leads to a velocity autocorrelation function that is also given by
an exponential, {\it i.e.}, $\langle \vec{v}(\tau) \cdot \vec{v}(0) \rangle = d v_0^2 \, e^{-\tau/\tau_{p}}$. 
	In addition, one can obtain the time-dependent diffusion coefficient $D(\tau)$ from the velocity autocorrelation function~\cite{doibook} 
%as $D(\tau) = d^{-1} \int_0^{\tau} \langle \vec{v}(t') \cdot \vec{v}(0) \rangle \, dt'$,
which, for the PRW model, is given by the following expression
\begin{equation}
D(\tau) 
=  d^{-1} \int_0^{\tau} \langle \vec{v}(t') \cdot \vec{v}(0) \rangle \, dt'
=v_0^2 \tau_{p} \left[ 1 -  \exp \left(-\frac{\tau}{\tau_{p}} \right) \right]~~.
\label{PRW-Dt}
\end{equation}
	Also, one can use the above result for $D(\tau)$ to obtain the MSD of the PRW model by integrating Eq.~(\ref{PRW-Dt}), that is,
\begin{equation} 
\langle \Delta r^2(\tau) \rangle = (2d) \int_0^{\tau} D(t') \, dt' = 2d v_0^2 \tau_{p} \left[ \tau - \tau_{p} \left(1-  e^{-\tau/\tau_{p}} \right) \right]~~.
\label{PRW-MSD}
\end{equation}
	The short-time behaviour ($\tau \ll \tau_{p}$) of the MSD can be easily obtained by expanding the exponential term in Eq.~(\ref{PRW-MSD}) up to second order, thus 
$\langle \Delta r^2(\tau) \rangle \approx d v_{0}^2 \tau^2$, which corresponds to the {\it ballistic regime}.
%\begin{equation}
% \langle (\Delta r)^2 (\tau) \rangle=4 \, \beta^2 \,\tau^2 ~~.
%\end{equation}
	On the other hand, for sufficiently long times, {\it i.e.}, $\tau \gg \tau_{p}$, one finds that the MSD should present a linear behaviour, that is,
$\langle \Delta r^2(\tau) \rangle \approx  2 d v_{0}^2 \tau_{p} (\tau - \tau_{p})$, and that corresponds to the  {\it normal diffusion regime}.
%\begin{equation}
% \langle (\Delta r)^2 (\tau)\rangle=8\beta^2\tau_{p}(\tau-\tau_{p}) ~~.
%\end{equation}
	We note that, although such limits are the same of those that can be observed for 
the experimentally obtained MSD presented in Fig.~\ref{exp_data_isolated_cell}, expression~(\ref{PRW-MSD}) 
does not fit the data very well.

	Although the difference between the experimental data and the results obtained from the PRW model can be already appreciated in the Fig.~\ref{exp_data_isolated_cell}, the effects of the variability of the cells are better realized from the probability
%\footnote{The position distribution $p(x)$ is also known as the van Hove distribution.} 
$p(x)$ of finding cells at a position $x$ after a time $\tau$ given that they all started at the origin, $x=0\,\mu$m.
%FALAR SOBRE DISTRIBUTIONS
	If one consider that, despite of being non-markovian, the stochastic processes defined by Eq.~(\ref{eq1}) in the case of the PRW model are still gaussian processes, hence the position distribution should be given by
% gaussian functions as well, that is
\begin{equation}
%p(x\,;\tau) 
p(x) = \left( \frac{1}{ 2 \pi \langle \Delta x^2 (\tau) \rangle }\right)^{1/2}
\exp \left[ - \frac{x^2}{2 \langle \Delta x^2 (\tau) \rangle} \right]~~,
\label{p_de_x}
\end{equation}
where $\langle \Delta x^2 (\tau) \rangle = d^{-1} \langle \Delta r^2(\tau) \rangle$
with the MSD given by Eq.~(\ref{PRW-MSD}).
	Note that, by assuming that the random walk is isotropic, one has that the distribution
$p(y)$ for the $y$-component should display the same functional form of the above equation with $\langle \Delta y^2 (\tau) \rangle = d^{-1} \langle \Delta r^2(\tau) \rangle$.

%	Also, for a two-dimensional substrate ($d=2$) one have that the distribution of distances, $r=\sqrt{x^2 + y^2}$,
%should be given by the {\bf Rayleigh distribution [CITAR E VERIFICAR]}, that is,
%\begin{equation}
%%%%p(r\,;\tau) 
%p(r) = \left( \frac{1}{ 2 \pi \langle \Delta r^2 (\tau) \rangle } \right)^{1/2}
%r \exp \left[ - \frac{r^2}{2 \langle \Delta r^2 (\tau) \rangle} \right] ~~,
%\label{p_de_r}
%\end{equation}
%with $\langle \Delta r^2 (\tau) \rangle$ given by Eq.~(\ref{PRW-MSD}).

	In the following we illustrate how the above theoretical results for the PRW model can be obtained by 
considering numerical simulations that lead to non-markovian stochastic processes
with a well-defined persistence time $\tau_p$.

%	Since the  velocity is defined by $v_x = x/\tau$,  one have that
%\begin{equation}
%p(v_x;\tau) = \left( \frac{1}{ 2 \pi \langle \Delta x^2 (\tau) \rangle }\right)^{1/2}
%\exp \left[ - \frac{x^2}{2 \langle \Delta v^2 (\tau) \rangle} \right]
%\end{equation}
%where  $\langle \Delta v_x^2 (\tau) \rangle =  \langle \Delta x^2 (\tau) \rangle/\tau^2$

%\begin{equation}
%p(v_{\tau};\tau) = \left( \frac{1}{ 2 \pi \langle \Delta v_{\tau}^2 (\tau) \rangle } \right)^{1/2}
%v_{\tau} \exp \left[ - \frac{v_{\tau}^2}{2 \langle \Delta v_{\tau}^2 (\tau) \rangle} \right]
%\end{equation}
%with $\langle \Delta v_{\tau}^2 (\tau) \rangle = \langle \Delta r^2 (\tau) \rangle/\tau^2$, with the MSD given by Eq.~(\ref{PRW-MSD}).

%	Such behaviour can be probed by a more sensitive analysis that considers the time-dependent diffusion coefficient $D(\tau)$, which can be evaluated numerically from the MSD as
%\begin{equation}
%D(\tau)=\frac{1}{2d}\frac{d \langle \Delta r^2(\tau) \rangle}{d\tau}~~,
%\label{coefdifusao}
%\end{equation}
%where we take $d=2$ for two-dimensional trajectories.

%%%%%%%%%%%%%%%%%%%%%%%%%%%%%%%%%%%%%%%%%%%%%%%%%%%%%%%%%%%%

\subsection{Non-markovian stochastic simulations}
\label{stochastic_simulations}

	Numerically, the main quantity used to quantify the motility of the cells
is their the mean-squared displacement (MSD), which is defined as
\begin{equation}
\langle \Delta r^2(\tau) \rangle = \langle \left[ \vec{r}(\tau) - \vec{r}(0) \right]^2 \rangle ~~,
\label{deltaR2}
\end{equation}
where $\langle \dots \rangle$ denotes averages over the trajectories of $N$ independent cells.
	In order to obtain the position of the cell $\vec{r}(t_{k})$ at a time $t_{k}$
one need to consider Eq.~(\ref{eq1}), which can be discretized and integrated according 
to the Euler's approximation scheme, that is,
$\vec{r}(t_{k+1}) = \vec{r}(t_k) + (\vec{f}_s / \gamma) \Delta t$,
where $t_{k+1} = t_{k} + \Delta t$ with $t_0=0\,$min.
%so that the position of the particle at a time $t_{k+1}$ is given by
%that is,
%\begin{equation}
%\vec{r}(t_{i+1}) \approx \vec{r}(t_i) + \frac{d\vec{r_i}}{dt} \delta t.
%\vec{r}(t_{k+1}) = \vec{r}(t_k) + \frac{\vec{f}_s}{\gamma} \Delta t ~~.
%\label{euller1}
%\end{equation}
%      Thus, just as in the usual brownian dynamics simulations, one can relate the intensity of the force $f_s$ to an effective free diffusion coefficient $D_0$ (which is proportional to $\gamma^{-1}$), so that Eq.~(\ref{euller1}) is rewritten (for a two-dimensional walk) as 
	By considering that the motion of the cells is restricted to a two-dimensional substrate ({\it i.e.}, $d=2$), %and also that $v_0 = f_s/\gamma$, 
the implementation is similar to the usual Brownian dynamics algorithm discussed in Ref.~\cite{gillespie1993amjphys}, that is,
\begin{equation}
% \vec{r}_{k+1} = \vec{r}_k + \sqrt{2D_0 \Delta t} \left[ \xi_x(0,1)\hat{x} +  \xi_y(0,1)\hat{y} \right] ~~,
%\vec{r}_{k+1} = \vec{r}_k + \sqrt{v_0^2 \, \Delta t \,} \, \left[ \xi_x(0,1)\hat{x} +  \xi_y(0,1)\hat{y} \right] ~~,
\vec{r}(t_{k+1}) = \vec{r}(t_k) + \sqrt{v_0^2 \, \Delta t \,} \, \left[ \xi_x(t_{k})\hat{x} +  \xi_y(t_{k})\hat{y} \right] ~~,
\end{equation}
where we assume that the variables $\xi_x(t_{k})$ and $\xi_y(t_{k})$ are independent  
%from each other.
%with $\xi_i(0,1)$ being a 
gaussian-distributed random variables 
%gaussian variables 
with zero mean and variance equal to one.
%     The implementation follows from the usual brownian dynamics algorithms presented in Ref.~\cite{gillespie1993amjphys}, where the $i$-th component of the force $\vec{f}_s$ is replaced by 
%%%$f_s \text{N}(0,1) /\sqrt{\Delta t}$, 
%%%where $\text{N}(0,1)$ denotes a gaussian S
% $f_s \, \xi_i(0,1) /\sqrt{\Delta t}$, 
%with $\xi_i(0,1)$ being a gaussian 
%variable with zero mean and variance equals to one.
      In order to obtain time-correlated velocities (or forces), we consider that the above defined random variables are generated by
the following recursive scheme~\cite{Janke2007},
\begin{equation}
%\xi_{k+1} = \rho \, \xi_{k} + \sqrt{1-\rho^2\,} \, \xi_{k}^{\prime},
\xi_{i}^{\,}(t_{k+1}) = \Lambda \, \xi_{i}^{\,}(t_{k}) + \sqrt{1-\Lambda^2\,} \, \xi_{i}^{\prime}(t_{k}),
\end{equation}
where $\xi_{i}^{\prime}(t_{k})$ is a gaussian-distributed random variable with zero mean and 
variance equal to one, with $\xi_{i}^{\prime}(t_{0})=\xi_{i}^{\,}(t_{0})$, and $\Lambda$ is a parameter 
in the range $0 < \Lambda <1$.
      This procedure lead to random varying velocities with an autocorrelation
%\begin{equation}
$A(t_k)=\langle \vec{\xi}(t_k) \cdot \vec{\xi}(0) \rangle \approx e^{-t_k/\tau_{p}}$
%\label{autocorrtk}
%\end{equation}
that is determined by a persistence ({\it i.e.},~characteristic autocorrelation) time  
$\tau_{p}$, just like in the PRW model.~As 
	discussed in Ref.~\cite{Janke2007}, the value of $\tau_p$ can be related to the parameter $\Lambda$ as
%\begin{equation}
% A(k)=\langle e_0e_k \rangle=\rho^k\equiv \exp \left(-\frac{k}{\tau_{\text{exp}}}\right).
%\end{equation}
%Thus, the exponential autocorrelation time  $\tau_{\text{exp}}$ is:
\begin{equation}
 \tau_{p}=-\frac{1}{\ln \Lambda} ~~.
\label{taup_rho}
\end{equation}
%     (we assume that the forces in the $xy$ directions are independent from each other), 
%     The value  $(2\gamma k_BT)1/2$ is assigned to the constant $f_s$, where $k_B$ is the Boltzmann constant and $T$ is the absolute temperature.
	Importantly, although this procedure ensures that the internal forces of the cell present an exponential time correlation, it keeps their statistical properties, {\it i.e.}, the stochastic process will be still a gaussian process with zero mean so that the distribution $p(x)$ will be given by Eq.~(\ref{p_de_x}) at any time $\tau$ (data not shown).
%and $p(r)$ will be given, respectively, by Eqs.~(\ref{p_de_x}) and~(\ref{p_de_r}) at any time $\tau$.
%$\langle \vec{f}_s(t) \rangle=\vec{0}$ and 
%$\langle \vec{f}_s(t+t_0) \cdot \vec{f}_s(t_0) \rangle=2 \Gamma e^{-t/\tau_{p}}$. 
%The autocorrelation function is given by:

%      Thus, just as in the usual brownian dynamics simulations, one can relate the intensity  of the force $f_s$ to an effective free diffusion coefficient $D_0$ (which is proportional to $\gamma^{-1}$), so that Eq.~(\ref{euller1}) is rewritten (for a two-dimensional walk) as 
%\begin{equation}
%%% \vec{r}_{k+1} = \vec{r}_k + \sqrt{2D_0 \Delta t} \left[ \xi_x(0,1)\hat{x} +  \xi_y(0,1)\hat{y} \right] ~~,
%\vec{r}_{k+1} = \vec{r}_k + \sqrt{v_0^2 \, \Delta t \,} \, \left[ \xi_x(0,1)\hat{x} +  \xi_y(0,1)\hat{y} \right] ~~,
%\end{equation}
%where we assume that the variables $\xi_x$ and $\xi_y$ are independent from each other.

%     the movement of a particle on a two-dimensional ($d=2$) substrate, the random force follows a gaussian distribution but here we consider that  the forces are correlated in time.

%gaussian, isotropic, ...

%Defining the free diffusion coefficient $D_0\equiv \frac {k_BT} {\gamma}$ we obtain the equation governing the particle dynamics by rewriting Eq.~(\ref{eq2}) as:

%%%%%%%%%%%%%%%%%%%%%%%%%%%%%%%%%%%%%%%%%%%%%%%%%%%%%%%%%%%%%%%%%
%%%%%%%%%%%%% FIG. 2:  trajectories for different rho
%%%%%%%%%%%%%%%%%%%%%%%%%%%%%%%%%%%%%%%%%%%%%%%%%%%%%%%%%%%%%%%%%
\begin{figure*}[!t]
\centering
\includegraphics[width=0.82\textwidth]{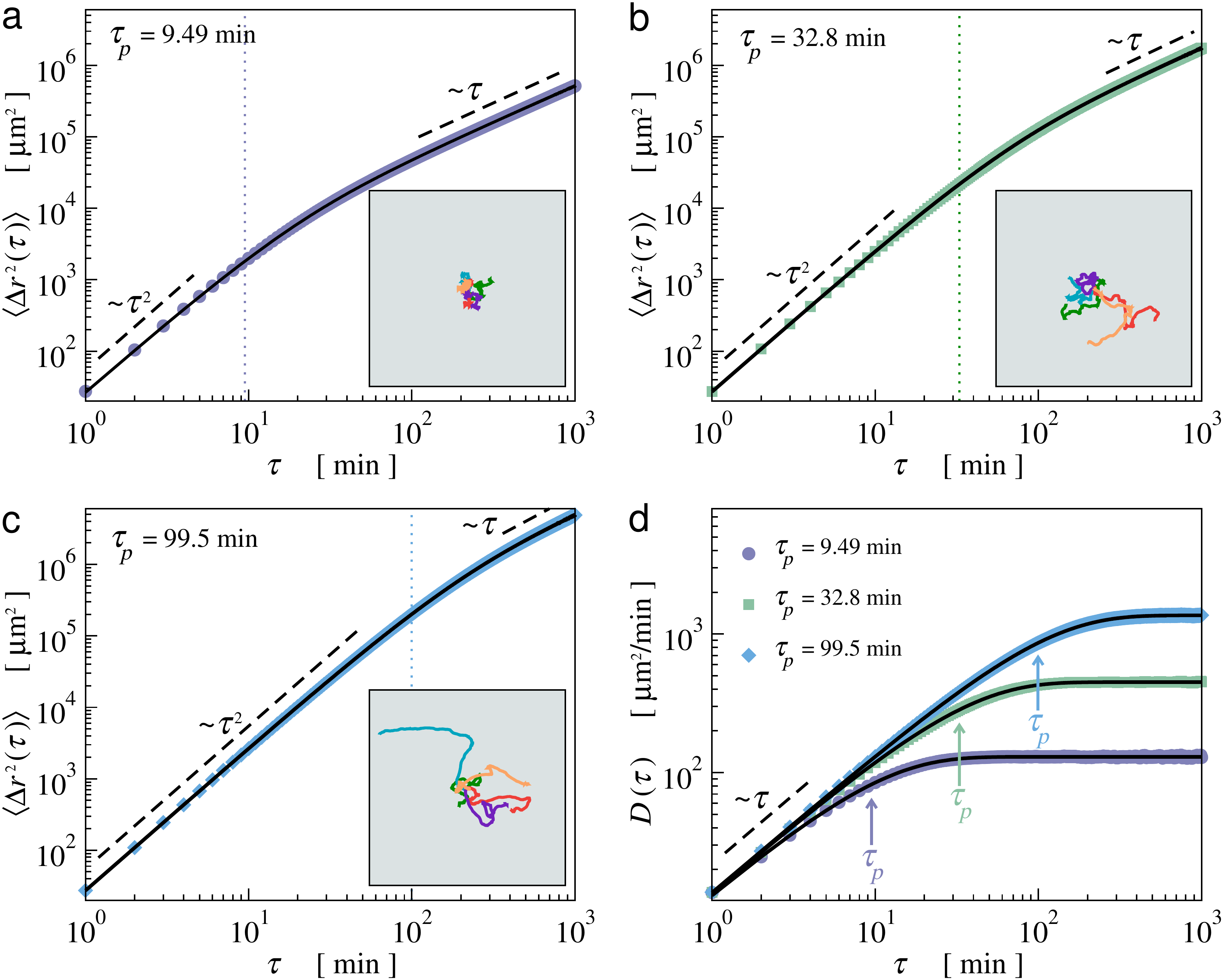}
\caption{(a)-(c)~Mean-squared displacement $\langle \Delta r^2 (\tau) \rangle$ obtained for the PRW model defined with different autocorrelation times $\tau_{p}$ determined by the parameter $\Lambda$: (a)~$\tau_{p}= 9.49\,$min ($\Lambda=0.9$), (b)~$\tau_{p}=32.8\,$min ($\Lambda=0.97$), and (c)~$\tau_{p}=99.5\,$min ($\Lambda=0.99$);
(d)~Time-dependent diffusion coefficient $D(\tau)$ for the three cases.
% 	Results presented here were obtained fromthe 
	 Filled symbols correspond to the numerical data obtained through non-markovian stochastic simulations considering a two-dimensional substrate ($d=2$), $v_0=3.7\,\mu \text{m}/\text{min}$, $\Delta t=1\,$min, and $N=10^6$ trajectories for each value of $\Lambda$.
%Results presented here were obtained from simulations using $\Delta t=1\,\text{min}$, $N=10^6$, and $D_0=1$ $\mu \text{m}^2/\text{min}$. 
%Continous and dot-dashed lines correspond to the data obtained through simulations, while dotted lines indicate non-linear power-law fits.
	 Straight black lines indicate 
the analytical results derived from the PRW model given by Eq.~(\ref{PRW-Dt}), $D(\tau)$, and Eq.~(\ref{PRW-MSD}), $\langle \Delta r^2 (\tau) \rangle$, while dashed lines denote the observed ballistic (for $\tau \ll \tau_p$) and normal diffusive (for $\tau \gg \tau_p$) regimes.
	Insets: typical two-dimensional trajectories obtained for the different autocorrelation times $\tau_p$.
%at $\tau=10^3\,$min.
}
%: (a)~$\tau_{p}\approx10\,$min~($\rho=0.9$), (b)~$\tau_{p}\approx33\,$min~($\rho=0.97$), and (c)~$\tau_{p}\approx99\,$min~($\rho=0.99$).}
%$D_0=1$ $\mu \text{m}^2/\text{min}$.}
\label{MSD_PRW_simulations}
\end{figure*}
%%%%%%%%%%%%%%%%%%%%%%%%%%%%%%%%%%%%%%%%%%%%%%%%%%%%%%%%%%%%%%%%%
%%%%%%%%%%%%%%%%%%%%%%%%%%%%%%%%%%%%%%%%%%%%%%%%%%%%%%%%%%%%%%%%%

      Figure~\ref{MSD_PRW_simulations} shows a comparison between the MSDs obtained from the above numerical scheme for different persistence times $\tau_{p}$ 
%by considering $\Delta t\,=\,1\,\text{min}$ and an effective 
%%%%diffusion coefficient $ D_0 = 1$ $\mu \text{m}^2/\text{min}$.
%velocity autocorrelation parameter $ v_0 = 3.7$ $\mu \text{m}/\text{min}$
and the theoretical results expected from the PRW model.
	As one can see in the inset panels, the trajectories  are more localized for shorter autocorrelation times $\tau_{p}$. 
	Higher values of the persistence times ($\tau_{p}=99.5\,$min) lead to more spreaded trajectories.
	Such diffusive-like behaviour, which is typical of the PRW model, can be probed by a more sensitive analysis that considers the time-dependent diffusion coefficient,
which can be evaluated numerically from the MSD as
\begin{equation}
D(\tau)=\frac{1}{2d}\frac{d \langle \Delta r^2(\tau) \rangle}{d\tau}~~.
\label{coefdifusao}
\end{equation}
	As displayed in Fig.~\ref{MSD_PRW_simulations}(d), the time-dependent diffusion coefficient $D(\tau)$ indicate that there is a clear separation between the ballistic regime at short times, where $D(\tau) \approx v_0^2 \tau$, and the normal diffusive-like behaviour at $\tau \gg \tau_p$, where
the diffusion coefficient is constant, {\it i.e.}, $D(\tau) \approx v_0^2 \tau_p$, just as expected from Eq.~(\ref{PRW-Dt}).

\section{Persistent random walk with variability (PRW-V)}
\label{PRW-Vmodel}

%\subsection{Persistent randowm walk with variability (PRW-V)}

	Now, in order to incorporate the cell-to-cell variability into our theoretical framework,
we assume that the distribution of persistence times is given by a
generalized inverse gamma distribution~\cite{crooksbook}, that is,
\begin{equation}
\rho(\lambda) = \frac{1}{\tau_p\Gamma(b)} \left( \frac{\tau_p}{\lambda} \right)^{b + 1} 
\exp \left( - \frac{\tau_p}{\lambda} \right) ~~,
\label{dist_lambda}
\end{equation}
where $\Gamma(b)$ is the usual gamma function.
	Here, it is convenient to consider 
the change of variables\footnote{Note that this change of variables 
should include the Jacobian factor, that is,
$\rho(\theta) = \left. \rho(\lambda) \right|_{\lambda=\tau_{p}/\theta} (\partial \lambda/\partial \theta)$.}
 $\theta=\tau_{p}/\lambda$, so that
$\rho(\theta) = \theta^{\,b-1}\,e^{-\theta} / \Gamma(b)$
and $e^{-\tau/\lambda} = e^{-\theta (\tau/\tau_{p})}$ in Eq.~(\ref{autocorr_general}).
	Thus, the distribution $\rho(\theta)$ is a simple gamma distribution~\cite{crooksbook} that can be used to evaluate both the mean persistence time, $\bar{\lambda}= \tau_p/(b-1)$,
%\begin{equation}
%\bar{\lambda} = \frac{\tau_{p}}{\alpha-1}
%\label{mean_persistence_time}
%\end{equation}
and the autocorrelation function, which is given by $A(\tau) = [ 1 + (\tau / \tau_{p} ) ]^{-b}$.
	Importantly, the expression~(\ref{dist_lambda}) is defined as a generic distribution that could be used to describe the experimental data, {\it e.g.}, the distribution of dwell times in Ref.~\cite{broedersz2020jrsocinter}. % (citar: cell to cell variability)
	Although several values can be attributed to the exponent $b$, here we consider that $b=2$ in order to have the mean value $\bar{\lambda}$ consistent with a given persistence time, that is, $\bar{\lambda}=\tau_p$.
	Hence, Eq.~(\ref{Vautocorr}) leads to a velocity autocorrelation function given by
\begin{equation}
\langle \vec{v}(\tau) \cdot \vec{v}(0) \rangle = d v_0^2 \, \left[ 1 + \left( \frac{\tau}{\tau_{p}} \right) \right]^{-2}~~.
\label{veloc-PRW-V-autocorr}
\end{equation}
	As a result, the time-dependent diffusion coefficient $D(\tau)$ and the MSD $\langle \Delta r^2(\tau) \rangle$ of the PRW-V model can be obtained by integration of the above expression (see Section~\ref{PRWmodel}), which yields, respectively,
%Time-correlation given by $A(\tau) = e^{-\tau/\lambda} = e^{-\theta (\tau/\tau_{p})}$
%with $\theta=\tau_{p}/\lambda$, where $\rho(\theta) = \frac{\theta^{\alpha-1}\,e^{-\theta}}{\Gamma(\alpha)}$
%time-dependent diffusion coefficient
\begin{equation}
D(\tau) 
%= \int_0^{\tau} \langle v(t') v(0) \rangle \, dt' 
= v_0^2 \tau_{p} \left\{ 1 -  \left[ 1 + \left( \frac{\tau}{\tau_{p}} \right) \right]^{-1} \right\}~~,
\label{PRW-V-Dt}
\end{equation}
and
%mean-squared displacement
\begin{equation}
\langle \Delta r^2(\tau) \rangle 
%= (2d) \int_0^{\tau} D(t') \, dt' 
= 2d v_0^2 \tau_{p} \left\{ \tau - \tau_{p} 
\ln \left[ 1 + \left( \frac{\tau}{\tau_{p}} \right) \right] \right\}~~.
\label{PRW-V-MSD}
\end{equation}
	Interestingly, the short and the later time behaviours of the time-dependent
diffusion coefficient, Eq.~(\ref{PRW-V-Dt}),
are similar to the results obtained from the PRW model in Section~\ref{PRWmodel}, that is, 
$D(\tau) \approx v_0^2 \tau $
for $\tau \ll \tau_p$,
and 
$D(\tau) \approx v_0^2 \tau_p$
for $\tau \gg \tau_p$.
	However, 
although the short time limit of Eq.~(\ref{PRW-V-MSD}) is similar to the MSD of the PRW model, {\it i.e.},
$\langle \Delta r^2(\tau) \rangle \approx d v_0^2 \tau^{2} $, the presence of
the logarithm leads to a slighly different behaviour for later times.
%$\langle \Delta r^2(\tau) \rangle \approx $

\begin{figure*}[!t]
\centering
\includegraphics[width=0.99\textwidth]{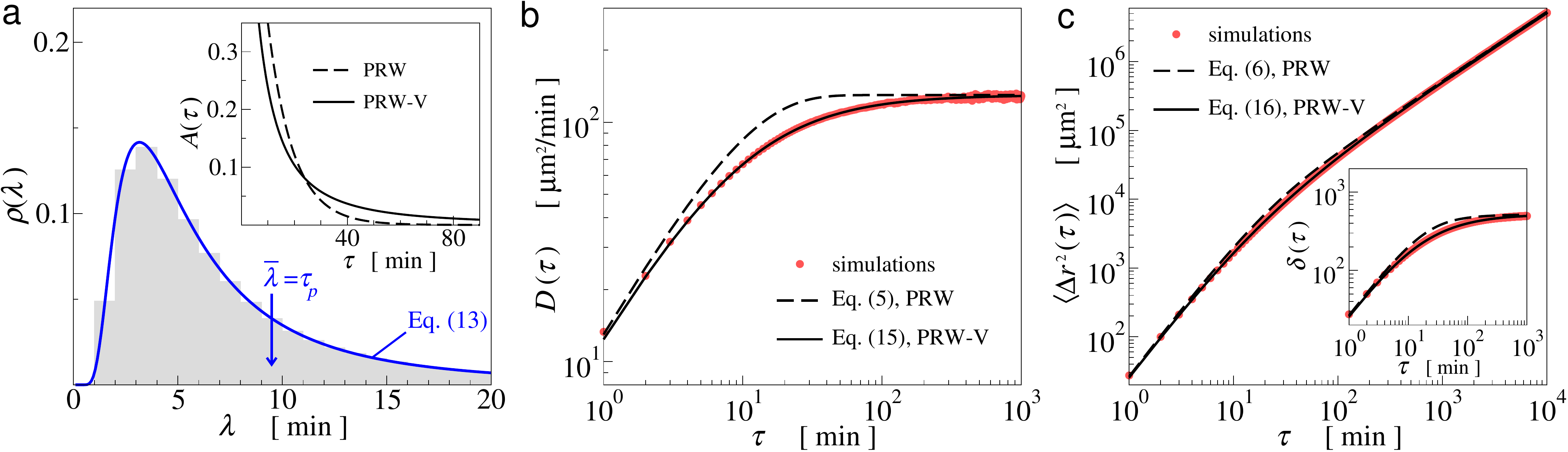}
\caption{
Numerical and theoretical results obtained for the PRW-V model considering that the persistence times are randomly distributed according to Eq.~(\ref{dist_lambda}) with %$b=2$ and 
$\bar{\lambda}=\tau_p=9.49\,$min ({\it i.e.},~$b=2$), $v_0=3.7\,\mu$m/min, $\Delta t=1\,$min, and $N=10^6$ independent two-dimensional ($d=2$) trajectories.
(a)~Numerically obtained histograms (grey bars) and the expected analytical distribution (straight blue line) of persistence times $\rho(\lambda)$. Inset
panel displays a comparison between the autocorrelation functions $A(\tau)$ of the PRW and the PRW-V models.
(b)~Time-dependent diffusion coefficient $D(\tau)$. (c)~Mean-squared displacement $\langle \Delta r^2(\tau) \rangle$
and the ratio $\delta(\tau)=\langle \Delta r^2(\tau) \rangle/\tau$ (inset panel).
%(a). (b). (c)~Mean squared displacement and Inset. Diffusion coefficient $D(\tau)$ for persistence times $\tau_{p}$ determined by different parameters $\rho$: $\rho=0.9$ ($\tau_{p}\approx 10\,$min), $\rho=0.97$ ($\tau_{p}=33\,$min), and $\rho=0.99$ ($\tau_{p}=99\,$min). Symbols are data obtained from simulations using $\Delta t=1\,\text{min}$, $N=10^6$, and $D_0=1$ $\mu\text{m}^2/\text{min}$. Straight lines are non-linear fits to Eq.~(\ref{PRW-Dt}).
}
\label{numresults_PRW-V}
\end{figure*}

	In Fig.~\ref{numresults_PRW-V} we include results for the PRW-V model and its comparison to the usual PRW model.
	The numerical simulations were carried out just as described in Section~\ref{stochastic_simulations}, however,
the $N$ trajectories were obtained by considering different values for $\Lambda$ in Eq.~(\ref{taup_rho}).
	In particular, we consider that the distribution of persistence times, $\lambda=-1/\ln \Lambda$, follows Eq.~(\ref{dist_lambda}) with $b=2$ and $\bar{\lambda}=\tau_p=9.49\,$min, as shown in Fig.~\ref{numresults_PRW-V}(a).
	Accordingly, the MSD presented in Fig.~\ref{numresults_PRW-V}(c) shows a behaviour that is
similar to the experimental results presented in Fig.~\ref{exp_data_isolated_cell}.
	Although the difference between the PRW and PRW-V seems to be small for the MSD and the ratio $\delta(\tau)$,
it is clear from the time-dependent diffusion coefficient $D(\tau)$ in Fig.~\ref{numresults_PRW-V}(b) that
the numerical results obtained from simulations with variability are better described by the PRW-V model.

%%%%%%%%%%
%% FIG. 3:  
%%%%%%%%%%

	Next we show that the differences between the usual PRW and the PRW-V models are significantly greater when one looks to the position and velocity distributions at later times.

\section{Distributions for the PRW-V model}
\label{PRW-Vdist}

\begin{figure*}[!t]
\centering
\includegraphics[width=0.95\textwidth]{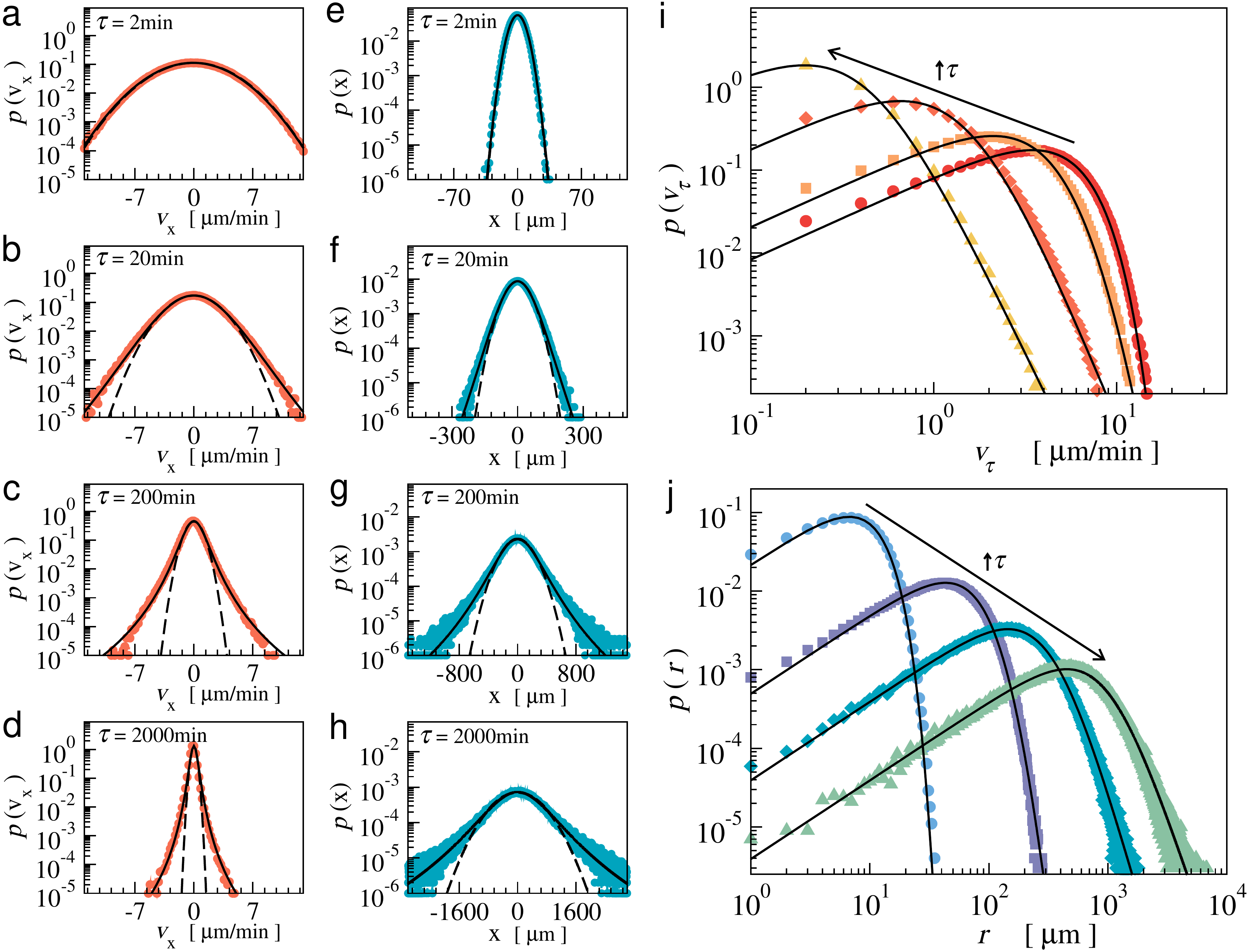}
\caption{Distributions of velocities and displacements obtained for the PRW-V defined by autocorrelation times determined by the distribution $\rho(\lambda)$ given by Eq.~(\ref{dist_lambda}) with $\bar{\lambda} = \tau_{p} = 9.49\,$min at different times $\tau$.
%\,min, $\tau=20$\,min, $\tau=200$\,min, and $\tau=2000$\,min,
	Filled symbols correspond to the histograms evaluated from the numerical data obtained by Brownian dynamics simulations with $v_0=3.7\,\mu$m/min, $\Delta t=1\,$min, and $N=10^6$ independent trajectories in a two-dimensional substrate ($d=2$).
%The histograms were evaluated with bin size $\Delta v=0.05\,\mu$m/min. Dashed lines denote the fits to maxwellian velocity distributions, Eq.~(\ref{max_dist}).
	Straight black lines in panels (a)-(h) correspond to component distributions $p(z)$ given by Eq.~(\ref{pearson_dist}), while the distributions $w(r)$ and $w(v_{\tau})$ in (i) and (j) are given by expression~(\ref{burr_dist}).
	Dashed black lines in panels (a)-(h) correspond to gaussian distributions, {\it e.g.}, Eqs.~(\ref{p_de_x}) and~(\ref{p_de_vx}).
%, while the distributions presented in (i) and (j) are given by Rayleigh distributions, i.e.,~Eqs.~(\ref{rayleigh_dists}).
	At later times, $\tau \gg \tau_p$, the distributions are well described by
the approximated expressions given by Eqs.~(\ref{p_de_z_approx}) and (\ref{w_de_z_approx}).
}
\label{histograms}
\end{figure*}

	Besides the position distribution $p(x)$, an often computed quantity from experiments is the distribution $p(v_x)$, which is related to the components of the velocity of the cells that can be defined as
$\vec{v}_{\tau} 
= [\vec{r}(\tau+t_0) - \vec{r}(t_0)]/\tau 
= \vec{r}(\tau)/\tau$.~If 
	the stochastic processes that lead to the motion of the cells are gaussian processes, the distributions of the components of $\vec{r}$ are given by Eq.~(\ref{p_de_x}), thus the distributions of the components $v_x$ of the velocities $\vec{v}_{\tau}$ will be also given by  gaussian distributions.~Thus, for the PRW model, in particular, the distribution $p(v_x)$ can be evaluated directly from Eq.~(\ref{p_de_x}) %and~(\ref{p_de_r}) and are given by
and is given by
%%	Since the velocity is defined by $v_x = x/\tau$,  one have that
\begin{equation}
%p(v_x;\tau)
p(v_x) = \left( \frac{1}{ 2 \pi \langle \Delta v_x^2 (\tau) \rangle }\right)^{1/2}
\exp \left[ - \frac{v_x^2}{2 \langle \Delta v_x^2 (\tau) \rangle} \right]~~,
\label{p_de_vx}
\end{equation}
%and 
%\begin{equation}
%%%%p(v_{\tau};\tau) 
%p(v_{\tau}) = \left( \frac{1}{ 2 \pi \langle \Delta v_{\tau}^2 (\tau) \rangle } \right)^{1/2}
%v_{\tau} \exp \left[ - \frac{v_{\tau}^2}{2 \langle \Delta v_{\tau}^2 (\tau) \rangle} \right]
%\label{p_de_vtau}
%\end{equation}
where  $\langle \Delta v_x^2 (\tau) \rangle =  \langle \Delta x^2 (\tau) \rangle/\tau^2$.
%and $\langle \Delta v_{\tau}^2 (\tau) \rangle = \langle \Delta r^2 (\tau) \rangle/\tau^2$,
%with the MSD given by Eq.~(\ref{PRW-MSD}).
	As shown in Figs.~\ref{histograms}(a)-(h), the histograms evaluated from the numerical data (filled symbols) obtained for the PRW-V model display a clear departure from the corresponding gaussian distributions, {\it i.e.}, Eqs.~(\ref{p_de_x}) and~(\ref{p_de_vx}), of the usual PRW model. % (denoted by dashed lines).
%	In fact, as indicated by the numerical data presented in Figs.~\ref{histograms}(i) and (h), gaussian distributions, or equivalently, Rayleigh distributions, i.e., $p(r)$ and $p(v_{\tau})$ given by Eqs.~(\ref{p_de_r}) and~(\ref{p_de_vtau}), respectively, are observed only for short times, i.e., $\tau \ll \tau_p=9.49\,$min.

%%%%%%%%%%
%% FIG. 4:  histograms 
%%%%%%%%%%

%\appendix
 
%\section{Distributions for the PRW-V model}

%By allowing that the persistence time $\lambda=\tau_p/\theta$ have a distribution Eq., 

	In order to obtain analytical expressions for the distributions $p(x)$ and $p(v_x)$ for the PRW-V model, 
one can consider that the gaussian distributions of the PRW model given by Eqs.~(\ref{p_de_x}) and~(\ref{p_de_vx}) are equivalent to conditional probabilities, that is, $p(z|\lambda)$, since
their variance $\langle \Delta z^2_{\lambda}(\tau) \rangle$
%$\langle \Delta x^2_{\lambda}(\tau) \rangle = 2 v_0^2 \lambda [\tau - \tau_p (1 - \exp(-\tau/\tau_p)) ]$.
depend on the persistence times $\lambda$ (here $z$ could denote either the component of the position $x$ or of the velocity $v_x$).
	In fact, those probabilities can be also written in terms of $\theta=\tau_p/\lambda$, that is, 
$p(z|\theta) = 1/\left( 2 \pi \langle \Delta z^2_{\theta} (\tau) \rangle \right)^{1/2}
\exp \left[ - z^2/ 2 \langle \Delta z^2_{\theta} (\tau) \rangle \right]$.
%\begin{equation}
%%%p(x\,;\tau) 
%p(z|\theta) = \left( \frac{1}{ 2 \pi \langle \Delta z^2_{\theta} (\tau) \rangle }\right)^{1/2}
%\exp \left[ - \frac{z^2}{2 \langle \Delta z^2_{\theta} (\tau) \rangle} \right]~~.
%\label{p_de_z_theta}
%\end{equation}
%%$\langle \Delta x^2_{\theta}(\tau) \rangle \approx (2 v_0^2 \tau_p \tau )/ \theta$.
%%with 
%%$\langle \Delta x^2_{\theta}(\tau) \rangle= 2 v_0^2 \lambda [\tau - \lambda (1 - \exp(-\tau/\lambda)) ]$
%%or 
%%$\langle \Delta x^2_{\theta}(\tau) \rangle = 2 v_0^2 ( \tau_p/\theta)^2 [\theta \, (\tau/\tau_p) - (1 - \exp(-\theta \, \tau/\tau_p)) ]$.
	Unfortunately, the expressions for the variances, {\it i.e.}, $\langle \Delta x^2_{\theta}(\tau) \rangle$ and
$\langle (\Delta v_x)^2_{\theta}(\tau) \rangle$,
may lead to non-trivial %distributions
joint distributions $p(z,\theta) = p(z|\theta)\, \rho(\theta)$
 that cannot be easily 
%integrated.
marginalized~\cite{jaynes}.
	Even so, at later times, $\tau \gg \tau_p$, the variances can be approximated
%the variances %of the PRW model 
by $\langle \Delta z^2_{\theta}(\tau) \rangle \approx \sigma_z^2/ \theta$,
with $\sigma_x^2 \approx 2 v_0^2 \tau_p \tau$ and $\sigma_{v_x}^2 = \sigma_x^2/\tau^2 \approx 2 v_0^2 \tau_p / \tau$.
%%% https://www.youtube.com/watch?v=8fYN_yKlAL0 NORMAL DISTRIBUTION WITH GAMMA PRIOR
%	Hence, one have that
	Thus, one can use %Eq.~(\ref{p_de_z_theta}) 
$p(z|\theta)$ together with the prior distribution $\rho(\theta)$ given in Section~\ref{PRW-Vmodel} to evaluate the distributions of position and velocity of the PRW-V at later times, {\it i.e.}, $\tau \gg \tau_p$, as
\begin{equation}
p(z) = \int_{0}^{\infty} p(z|\theta) \, \rho(\theta) \, d\theta
%\approx
%\left( \frac{1}{ 2 \pi \sigma^2  }\right)^{1/2}
%\frac{3 \pi^{1/2}}{4 \left( 1 + \frac{x^2}{2 \sigma^2 }  \right)^{5/2} }
%~~,
\approx
\frac{3}{4  \sqrt{2 \sigma_z^2 }} \left( 1 + \frac{z^2}{2 \sigma_z^2}  \right)^{-5/2}~~.
\label{p_de_z_approx}
\end{equation}
where $z$ can be either the component of the position $x$ or of the velocity $v_x$,
and $\sigma_z^2$ denote estimates for the corresponding variances, 
{\it i.e.}, $\langle \Delta x^2(\tau) \rangle$ and
$\langle \Delta v_x^2(\tau) \rangle$.
%, which are given in terms of the MSD of the PRW model as in Section~\ref{PRWmodel}, i.e., Eq.~(\ref{PRW-MSD}) with the same $v_0$ and $\tau_p$.
	Indeed, by considering $v_0=3.7\,\mu$m/min,  $\tau_p=9.49\,$min, and $\tau=2000\,$min,
one gets 
$\sigma_x^2 \approx 2 v_0^2 \tau_p \tau \approx 5.2 \times 10^5\,\mu$m$^2$, so that the 
%$2\sigma_x^2 \approx 4 v_0^2 \tau_p \tau = 1.039 \times 10^6\,\mu$m$^2$, so that the 
agreement of the above expression, Eq.~(\ref{p_de_z_approx}), for $p(x)$ and the numerical data presented for the PRW-V in 
Fig.~\ref{histograms}(h) is remarkable.
	A good agreement is also observed between the numerical results and the $p(v_x)$ given by Eq.~(\ref{p_de_z_approx}) with 
%$2\sigma_{v_x}^2 \approx 4 v_0^2 \tau_p / \tau = 0.2598 \mu$m$^2$/s$^2$.
%$\sigma_{v_x}^2 = \sigma_{x}^2/\tau^2 \approx 2 v_0^2 \tau_p / \tau \approx 0.259\,\mu$m$^2$/s$^2$
$\sigma_{v_x}^2 = \sigma_{x}^2/\tau^2 \approx 2 v_0^2 \tau_p / \tau \approx 0.13\,\mu$m$^2$/min$^2$
in Fig.~\ref{histograms}(d), {\it i.e.},~at later times.

	It is worth noting that the expression~(\ref{p_de_z_approx}) is a particular case of a more general distribution called 
Pearson (type VII) distribution~\cite{crooksbook}, which is given by % page 60 crooks
%``generalized'' dist corresponds to the Pearson VII distribution page 66 crooks
\begin{equation}
p(z) = \frac{1}{|s| \, B(m-1/2,1/2) } \left( 1 +  \frac{z^2}{s^2}  \right)^{-m}
\label{pearson_dist}
\end{equation}
where $B(l,c)= \int_{0}^{1} u^{l-1} (1-u)^{c-1} du$ is the so-called beta function.
%Relativistic BreitWigner crooks (9.8) page 64
	Interestingly, one can assume that 
$s^2= (2m -3) \sigma_z^2$ (for $m>3/2$),
so that $p(z)$ yields exactly the same distribution given by Eq.~(\ref{p_de_z_approx}) at later times,
{\it i.e.}, with $m=5/2$, $B(2,1/2)=4/3$, and $s^2=2 \sigma_z^2$.
	In fact, as shown in Fig.~\ref{histograms}(a)-(h), different values of the exponent
$m$ can be used to interpolate between
the later times ($\tau \gg \tau_p$), where the distributions $p(z)$ are given by Eq.~(\ref{p_de_z_approx}) with 
$m=5/2$, 
and short times ($\tau \ll \tau_p$), 
where the exponent 
$m$ assumes large values
 and 
the distributions $p(z)$ given by Eq.~(\ref{pearson_dist}) correpond to
gaussian distributions\footnote{One can verify that by
considering the relationship $B(l,c) =\Gamma(l) \, \Gamma(c)/\Gamma(l+c)$, and replace
$B(m-1/2,1/2)$ and $s= \sqrt{2(m-3/2)\sigma^2_z\,}$ into Eq.~(\ref{pearson_dist}),
so that the limit of $m \rightarrow \infty$ of the Pearson distribution
yields the prefactor $1/\sqrt{2\pi \sigma_z^2 \,}$ times
a $z$-dependent exponential, {\it i.e.},~$\exp(-z^2/2\sigma^2_z)$,  just like the gaussian distributions 
given by Eqs.~(\ref{p_de_x}) and.~(\ref{p_de_vx}).}
 with variance $\sigma_z^2$.% given by the PRW model, i.e., Eqs.~(\ref{p_de_x}) and~(\ref{p_de_vx}) of 

%===============

%LIMIT FOR LARGE $m$, $B(l,q) \Gamma(l+q)=\Gamma(l) \Gamma(q)$ so that
%\begin{equation}
%\frac{1}{B(m-1/2,1/2)} = \frac{\Gamma(m)}{\Gamma(m-1/2) \Gamma(1/2)}
%\end{equation}
%where $\Gamma(1/2)=\sqrt{\pi}$.
%	Then, the limit is
%\begin{equation}
%\lim_{m \rightarrow \infty} \frac{1}{\sqrt{2(m-3/2)} \, B(m-1/2,1/2)} = \frac{1}{\sqrt{2\pi}} \lim_{m \rightarrow \infty} \frac{\Gamma(m)}{\sqrt{m-3/2} \, \Gamma(m-1/2)} \approx
%\frac{1}{\sqrt{2\pi}}
%\end{equation}

%===============

	Next, we turn our attention to the distributions of distances, $r=\sqrt{x^2 + y^2}$, and velocities, $v_{\tau}=r/\tau$, as they are commonly evaluated by the experimentalists.
	For the two-dimensional ($d=2$) PRW model one have that those distributions are given by the 
Rayleigh distributions~\cite{rayleigh}, 
% "The Problem of the Random Walk", Nature 1905 vol.72 p.318
that is,
\begin{equation}
%%p(r\,;\tau) 
w(r) = 
 \frac{2\,r}{\langle \Delta r^2 (\tau) \rangle } 
%\left( \frac{1}{ 2 \pi \langle \Delta r^2 (\tau) \rangle } \right)^{1/2}
\exp \left[ - \frac{r^2}{\langle \Delta r^2 (\tau) \rangle} \right] ~~
%\label{p_de_r}
~~~~~~~~
\text{and}
~~~~~~~~
w(v_{\tau}) 
=  \frac{2\,v_{\tau}}{ \langle \Delta v_{\tau}^2 (\tau) \rangle } 
%= \left( \frac{1}{ 2 \pi \langle \Delta v_{\tau}^2 (\tau) \rangle } \right)^{1/2}
 \exp \left[ - \frac{v_{\tau}^2}{\langle \Delta v_{\tau}^2 (\tau) \rangle} \right]~~,
%\label{p_de_vtau}
\label{rayleigh_dists}
\end{equation}
where $\langle \Delta v_{\tau}^2 (\tau) \rangle = \langle \Delta r^2 (\tau) \rangle/\tau^2$,
with the MSD of the PRW model $\langle \Delta r^2 (\tau) \rangle$ given by Eq.~(\ref{PRW-MSD}).
%%%%%%%%%%%%%%%%%%%%% w(r) PRW %%%%%%%%%%%%%%%%%%%%%
%y = 2*x*exp(-x^2/(102.215))/102.215
%y = 2*x*exp(-x^2/(6061.1728))/6061.1728
%y = 2*x*exp(-x^2/(99002.7889))/99002.7889
%y = 2*x*exp(-x^2/(1034413.109))/1034413.109
%%%%%%%%%%%%%%%%%%%%% w(r) PRW-V later times %%%%%%%%%%%%%%%%%%%%%
%y = 4*x/( (1034413.109)   * ( 1 + x^2/(1034413.109)  )^3 )  OK
%	In addition, one can use the above expression, Eq.~(\ref{p_de_r}), to evaluate the distribution of velocities for the PRW model by considering $v_{\tau}=r/\tau$, and that yields
%\begin{equation}
%%%%p(v_{\tau};\tau) 
%w(v_{\tau}) 
%=  \frac{2}{ \langle \Delta v_{\tau}^2 (\tau) \rangle } 
%%%%= \left( \frac{1}{ 2 \pi \langle \Delta v_{\tau}^2 (\tau) \rangle } \right)^{1/2}
%v_{\tau} \exp \left[ - \frac{v_{\tau}^2}{\langle \Delta v_{\tau}^2 (\tau) \rangle} \right]
%\label{p_de_vtau}
%\end{equation}
%with $\langle \Delta v_{\tau}^2 (\tau) \rangle = \langle \Delta r^2 (\tau) \rangle/\tau^2$,
%with the MSD given by Eq.~(\ref{PRW-MSD}).
%%%%%%%%%%%%%%%%%%%%% w(v_tau) PRW %%%%%%%%%%%%%%%%%%%%%
%y = 2*x*exp(-x^2/(25.55))/25.55
%y = 2*x*exp(-x^2/(15.15))/15.15
%y = 2*x*exp(-x^2/(2.475))/2.475
%y = 2*x*exp(-x^2/(0.2586))/0.2586
%%%%%%%%%%%%%%%%%%%%% w(r) PRW-V later times %%%%%%%%%%%%%%%%%%%%%
%y = 4*x/( (0.2586)   * ( 1 + x^2/(0.2586)  )^3 )  
	The corresponding distributions for the PRW-V can be obtained by considering a similar approach
that lead to Eq.~(\ref{p_de_z_approx}), so that, at least for the limiting case where $\tau \gg \tau_p$, one have that
\begin{equation}
w(z) = \int_{0}^{\infty} w(z|\theta) \, \rho(\theta) \, d\theta
%\approx
%\left( \frac{1}{ 2 \pi \sigma^2  }\right)^{1/2}
%\frac{3 \pi^{1/2}}{4 \left( 1 + \frac{x^2}{2 \sigma^2 }  \right)^{5/2} }
%~~,
\approx
\frac{4 z}{\Delta^2_z } \left( 1 + \frac{z^2}{\Delta_z^2}  \right)^{-3}~~.
\label{w_de_z_approx}
\end{equation}
where $z$ can be either the distance $r$,
so that $\Delta^2_r \approx 2 d v_0^2 \tau \tau_p$,
or the velocity $v_{\tau}$,
with $\Delta^2_{v_{\tau}} = \Delta^2_r/\tau^2 \approx 2 d v_0^2 \tau_p/\tau$.
	Indeed, as shown in Figs.~\ref{histograms}(i) and (j), %(h)
 the above expression
is very accurate to describe the numerical data at later times, {\it i.e.}, $\tau =2000\,$min.
%and $\sigma_z^2$ denotes the corresponding variances, 
% i.e., $\langle \Delta x^2_{\theta}(\tau) \rangle$ and
%$\langle (\Delta v_x)^2_{\theta}(\tau) \rangle$, that are
%given by the PRW model as in Section~\ref{PRWmodel}.
	Interestingly, Eq.~(\ref{w_de_z_approx}) can be considered a limiting case of the
Burr (type XII) distribution~\cite{crooksbook}, which is given by
\begin{equation}
w(z) = \frac{2 n z}{s^2} 
\left(
1 + \frac{z^2}{s^2} 
\right)^{-(n+1)}~~,
\label{burr_dist}
\end{equation}
where one can assume the specialization $s^2=(n-1)\Delta_z^2$ (for $n>1$).~Accordingly, such 
	distribution approaches the Rayleigh distributions, Eqs.~(\ref{rayleigh_dists}), for large values of $n$, and
it is equal to Eq.~(\ref{w_de_z_approx}) if $n=2$, which is the limit for later times.
	As indicated by the numerical data presented in Figs.~\ref{histograms}(i) and~\ref{histograms}(j),
the above expression can be used to describe the numerical data obtained for the PRW-V remarkably well.
	Since both expressions~(\ref{p_de_z_approx}) and~(\ref{w_de_z_approx}) were obtained from the same distribution of 
persistence times $\rho(\lambda)$, one can expect that the exponents defined in the distributions~(\ref{pearson_dist}) and~(\ref{burr_dist}) should be related as $n=m-1/2$. %, with $n \rightarrow 5/2$ and $n \rightarrow 2$ at later times.
	Indeed, that relationship can be verified to be generally valid for all the data displayed in Fig.~\ref{histograms}.
	Gaussian-like distributions, 
%(with $m\approx 26.5$), 
or equivalently, Rayleigh-like distributions (with $n\approx 26$), % i.e., $p(r)$ and $p(v_{\tau})$ given by Eqs.~(\ref{p_de_r}) and~(\ref{p_de_vtau}), respectively, 
are observed only for short times, {\it i.e.}, $\tau = 2\,$min.
	Finally, it is worth mentioning that Eq.~(\ref{burr_dist}) is similar to the $q$-Weibull distribution
that was used to fit the experimental data obtained for
%	As shown by experiments and simulations, similar but 
the diffusive behaviour of non-isolated cells~\cite{podesta2017plosone}.
	A comparison between the two expressions lead to $q=(n+2)/(n+1) \approx 1.33$ at later times ({\it i.e.}, $n=2$ for $\tau \gg \tau_p$), which is close to the values found in Ref.~\cite{podesta2017plosone}, even though here we are 
considering only isolated cells.

\section{Concluding remarks}

%	As shown by experiments and simulations, similar but more complex diffusive behaviours can be observed for non-isolated cells~\cite{podesta2017plosone}.
%(*) Many models are used in the study of active diffusion, with the Vicsek model being one of the most used. Experimental and simulated results using the Vicsek model in the study of cell migration were obtained in Ref.~\cite{podesta2017plosone}.

	In this work we introduce novel computational and theoretical approaches in order to explore the effect of time-correlated forces in the diffusion of active specimens and apply them to describe cell motility.~By
	considering non-markovian stochastic simulations that display exponential autocorrelations with 
well-determined persistence times, %$\tau_{p}$ (see Eq.~\ref{taup_rho}), 
we were able to generate
%adapt the usual Brownian dynamic simulations in order to simulate 
persistent random walks that % has an analytical solution.
incorporate the %cell-to-cell 
biological variability 
that is inherent to the %biomolecular 
processes that occur in the intra-cellular media.
	Our numerical and analytical results suggest that the heterogeneity of 
persistence times can be directly related to the non-gaussianity of the distributions
observed in experiments. 
%	Importantly, our results are consistent with experimental results obtained for the motion of isolated cells~\cite{cox2008plosone}, and our approach might be used in simulations of cell motion in crowded environments~\cite{podesta2017plosone}.
	Although we have considered a specific distribution $\rho(\lambda)$, 
depending on the cell line it might be interesting to consider
slightly different values for the exponent $b$
in Eq.~(\ref{dist_lambda}).~Importantly, 
	here we have choose $b=2$ in order to set $\bar{\lambda}=\tau_p$, but 
reliable experimental estimates for that value might be inferred from the 
velocity autocorrelation function, Eq.~(\ref{veloc-PRW-V-autocorr}), since
%Eq.~(\ref{veloc-PRW-V-autocorr}), since 
$A(\tau)=[1 + (\tau/\tau_p)]^{-b}$ for the PRW-V model.
	Also, we believe that, with the ideas developed here, the generalization of the distributions $w(r)$ and $w(v_{\tau})$ to three-dimensional substrates should be straightforward.

%	Also, we believe that the approches presented hereshould be  to three-dimensional substrates.

	Finally, it is worth mentioning that subdiffusive motility, {\it i.e.}, with the MSD characterized by Eq.~(\ref{anomdif}) with $\alpha<1$, % and that are usually observed in ,
%APL BIOENGINEERING 2, 026112 (2018)
%Anomalously diffusing and persistently migrating cells in 2\uppercase{D} and 3\uppercase{D} culture environments
%Igor D. Luzhansky and Alyssa D. Schwartz and Joshua D. Cohen and John P. MacMunn and Lauren E. Barney and Lauren E. Jansen and Shelly R. Peyton
can be considered within 
the non-markovian stochastic simulations of the PRW-V model simply by adding a
hookean-like force, {\it i.e.}, $\vec{F}(\vec{r}\,)=-\kappa \vec{r}$, to the right hand side of Eq.~(\ref{eq1}).
	We think that could be an interesting way to extend the present framework
to cases where the cells are moving in confined and/or crowded environments.

\vspace{-0.1cm}

%%%%%%%%%%
%% FIG. 5: experimental data
%%%%%%%%%%
%\begin{figure}[!t]
%\centering
%%%\includegraphics[width=0.43\textwidth]{grafico_exp_comparison}
%\caption{Experimental data for the mean-squared displacement $\langle \Delta r^2 (\tau) \rangle$ of isolated cells taken from Ref.~\cite{cox2008plosone}. Inset: time-dependent diffusion coefficient $D(\tau) = \langle \Delta r^2 (\tau) \rangle / (2 d\tau)$.}
%\label{experimental_data}
%\end{figure}

%%%%%%%%%%%%%%%%%%%%%%
% Acknowledgements %%%
%%%%%%%%%%%%%%%%%%%%%%

\section*{Acknowledgements}

	The authors acknowledge the useful discussions with Professor Marcelo Lobato Martins and
%	T. N. Azevedo thanks CAPES (001) for the Scholarship.
%	L. G. Rizzi acknowledges
%Marcelo L. Martins for the insightful discussions about cell migration and
% Alvaro V. N. C. Teixeira for the useful discussions about the implementation of the Brownian dynamics simulations.
	the financial support of the Brazilian agencies, CAPES (code 001) for the Scholarship, CNPq (Grants 
N\textsuperscript{o}  306302/2018-7 
% Bolsas de Produtividade em Pesquisa - PQ
and 
N\textsuperscript{o} 426570/2018-9), and 
% Chamada MCTIC/CNPq N 28/2018 - Universal/Faixa A - Ate R$ 30.000,00
	FAPEMIG (Process APQ-02783-18).
%, although no funding was released until the submission of the present work.

%% \bibliography{biblio}

%%%%%%%%%%%%%%%%%%%%%

\section*{References}

%\bibliographystyle{elsarticle-num}

%\bibliography{biblio}{}

%\begin{thebibliography}{00}

%% \bibitem{label}
%% Text of bibliographic item

%\bibitem{}

%\end{thebibliography}

\end{document}